\documentclass[sigconf,authorversion]{acmart}

\AtBeginDocument{%
  \providecommand\BibTeX{{%
    \normalfont B\kern-0.5em{\scshape i\kern-0.25em b}\kern-0.8em\TeX}}}

\setcopyright{acmcopyright}
\copyrightyear{2018}
\acmYear{2018}
\acmDOI{10.1145/1122445.1122456}

\acmConference[SUI '21]{Virtual Conference}{November 09--10, 2021}{Virtual Conference}
\acmBooktitle{Woodstock '18: ACM Symposium on Neural Gaze Detection,
  June 03--05, 2018, Woodstock, NY}
\acmPrice{15.00}
\acmISBN{978-1-4503-XXXX-X/18/06}

\usepackage{colortbl} 
\usepackage{color,soul}
\sethlcolor{cyan}
\renewcommand\hl[1]{#1}

\begin{document}

\title{Accuracy Evaluation of Touch Tasks in Commodity Virtual and Augmented Reality Head-Mounted Displays}


\author{Daniel Schneider}
\email{daniel.schneider@hs-coburg.de}
\affiliation{%
    \institution{Coburg University of Applied Sciences and Arts}
    \city{Coburg}
  \country{Germany}
}

\author{Verena Biener}
\email{verena.biener@hs-coburg.de}
\affiliation{%
  \institution{Coburg University of Applied Sciences and Arts}
    \city{Coburg}
  \country{Germany}
}

\author{Alexander Otte}
\email{alexander.otte@hs-coburg.de}
\affiliation{%
  \institution{Coburg University of Applied Sciences and Arts}
    \city{Coburg}
  \country{Germany}
}

\author{Travis Gesslein}
\email{travis.gesslein@hs-coburg.de}
\affiliation{%
  \institution{Coburg University of Applied Sciences and Arts}
    \city{Coburg}
  \country{Germany}
}

\author{Philipp Gagel}
\email{philipp.gagel@stud.hs-coburg.de}
\affiliation{%
  \institution{Coburg University of Applied Sciences and Arts}
    \city{Coburg}
  \country{Germany}
}

\author{Cuauhtli Campos}
\email{cc.mijangos@gmail.com}
\affiliation{%
    \institution{University of Primorska}
    \city{Koper}
  \country{Slovenia}
}

\author{Klen Čopič Pucihar}
\email{klen.copic@famnit.upr.si}
\affiliation{%
  \institution{University of Primorska}
    \city{Koper}
  \country{Slovenia}
}

\author{Matjaž Kljun}
\email{matjaz.kljun@famnit.upr.si}
\affiliation{%
  \institution{University of Primorska}
    \city{Koper}
  \country{Slovenia}
}

\author{Eyal Ofek}
\email{eyalofek@microsoft.com}
\affiliation{%
    \institution{Microsoft Research}
    \city{Redmond, Washington}
  \country{United States}
}

\author{Michel Pahud}
\email{mpahud@microsoft.com}
\affiliation{%
  \institution{Microsoft Research}
    \city{Redmond, Washington}
  \country{United States}
}

\author{Per Ola Kristensson}
\email{pok21@cam.ac.uk}
\affiliation{%
    \institution{Department of Engineering, University of Cambridge}
    \city{Cambridge}
  \country{United Kingdom}
}

\author{Jens Grubert}
\email{jg@jensgrubert.de}
\affiliation{%
  \institution{Coburg University of Applied Sciences and Arts}
    \city{Coburg}
  \country{Germany}
}

\renewcommand{\shortauthors}{Schneider, et al.}

\begin{abstract}
  An increasing number of consumer-oriented head-mounted displays (HMD) for augmented and virtual reality (AR/VR) are capable of finger and hand tracking.  We report on the accuracy of off-the-shelf VR and AR HMDs when used for touch-based tasks such as pointing or drawing. Specifically, we report on the finger tracking accuracy of the VR head-mounted displays Oculus Quest, Vive Pro and the Leap Motion controller, when attached to a VR HMD, as well as the finger tracking accuracy of the AR head-mounted displays Microsoft HoloLens 2 and Magic Leap. 
We present the results of two experiments in which we compare the accuracy for absolute and relative pointing tasks using both human participants and a robot. 
The results suggest that HTC Vive has a lower spatial accuracy than the Oculus Quest and Leap Motion and that the Microsoft HoloLens 2 provides higher spatial accuracy than Magic Leap One. These findings can serve as decision support for researchers and practitioners in choosing which systems to use in the future.

\end{abstract}




\begin{teaserfigure}
  \includegraphics[width=\textwidth]{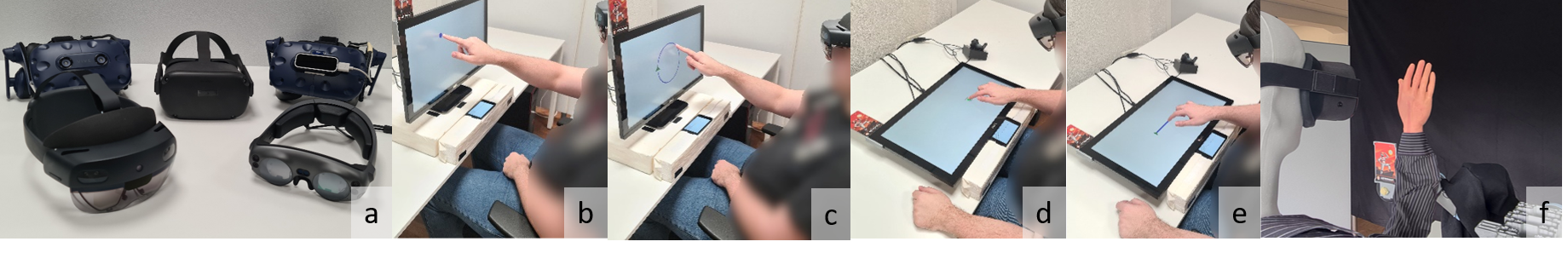}
  \caption{a: HMDs used for the conditions: HTC Vive Pro (top left), Oculus Quest (top center), Leap Motion (top right), Microsoft HoloLens 2 (bottom left), MagicLeap 1 (bottom right). Human participants study participant conducting in \textsc{Vertical} orientation b: \textsc{Target Acquisition} task,  c: \textsc{Shape Tracing } task and \textsc{Horizontal} orientation d: \textsc{Target Acquisition} task, e: \textsc{Shape Tracing} task, f: robot study setup.}
  \label{fig:teaser}
\end{teaserfigure}

\maketitle

\section{Introduction}
The ability to see one's hands and finger movement inside Virtual Reality (VR) opens up opportunities for natural interaction in VR. This raises the potential to bridge the virtual space with the limited input space available in today's touch surfaces such as laptops, desktops and touchscreen-based devices.

Lately, commercial VR head-mounted displays (HMDs) are progressing to 'inside-out' tracking using multiple built-in cameras \cite{han2020megatrack}. Inside-out tracking allows a simple setup of the VR system, and the ability to work in uninstrumented environments. Tracking the user hands in real-time is a potentially useful capability, that is already implemented in a number of consumer-oriented AR and VR HMDs. 
Given the capability of spatial hand and finger sensing, it is possible to see VR extending the input space of existing computing devices, such as touchscreens and keyboards, to a space around the devices, thereby enhancing their usage. A knowledge worker can use existing 2D surfaces  as well as using space around the devices and in front of the screen, reachable while sitting, to represent and manipulate additional information. In this scope, the applicability of hand and finger tracking has a crucial dependency on the accuracy of tracking the user's fingers: While detection of certain finger gestures can be achieved using only coarse positional accuracy, interaction with real physical objects requires accurate grounded positioning of the VR coordinate system relative to the relevant objects, and a good spatial positioning of the user's fingers. 
Further, when mixing the interaction with physical input devices and 3D hand tracking, there is special importance in the relationship between these. Different techniques of finger sensing, the ability to select an object, and the minimal distance between such an object and nearby ones to prevent selection errors, are elements that need to be considered. Also, hand tracking algorithms must cope with self-occlusions \cite{han2020megatrack} and refer to the estimation of the complete hand pose, including all finger joints, the hand position itself, or the position of an end effector, such as the fingertip. 
Within this paper, we focus on touch-based interactions, that is, tasks in which the user interacts with physical or digital surfaces, for example, for the purpose of selecting objects (such as buttons) or for drawing on surfaces \cite{romat2021flashpen}. While the achievable accuracy is dependent on multiple factors (e.g., the accuracy and latency of actual finger and hand tracking sensors and algorithms as well as the localization system of the respective HMD), we deliberately focus on the overall accuracy achievable with commodity off-the-shelf HMDs as this is of major concern for many practitioners and researchers who want or need to use the HMDs without modifying individual subsystems. Specifically, we report results from a controlled study with human participants ($n = 20$), investigating the accuracy of VR headsets (HTC Vive, Oculus Quest) and a LeapMotion sensor, as well as AR headsets (Microsoft Hololens 2 and Magic Leap 1). Further, we complemented the study with human participants with a study using a robotic-arm, which allows for better repeatability of measurements, at the cost of ecological validity. Our findings suggest that for VR HMDs, the HTC Vive results in significantly lower spatial accuracy (mean distance between a target and the sensed fingertip location of around 37 mm with a standard deviation of 20 mm) compared to both Oculus Quest (mean = 16 mm, sd = 9 mm) and Leap Motion  (mean = 13 mm, sd = 6 mm). For AR HMDs the Microsoft HoloLens 2 provides a significantly higher spatial accuracy  (mean distance between a target and the sensed fingertip location of around 15 mm with a standard deviation of 9 mm) compared to the Magic Leap One (m = 40 mm, sd = 16 mm).
We hope the reported results will support the design of 3D work spaces around touch devices in AR and VR. 

\section{Related Work}



Besides stylus-based \cite{pham2019pen, batmaz2020precision} and controller-based \cite{speicher2018selection, arora2017experimental} input, hand and finger input is important for 3D interaction techniques in XR for selection, spatial manipulation, navigation and system control \cite{laviola20173d}. There are recent surveys on these techniques~\cite{argelaguet2013survey, mendes2019survey}. 

The performance of the Leap Motion controller was already studied in various context.
Weichert et al. \cite{weichert2013analysis} evaluated accuracy and repeatability of the Leap Motion Controller using a pen (with variable diameters form 3 to 10mm) mounted on a robotic arm (position accuracy of 0.2mm). The authors indicated an accuracy of below 0.2 mm in a static measurement setup. However, the measured volume only encompassed 20~cm along each axis. This setup replicates one intended use case of the Leap Motion controller, namely lying flat on a desk with fingers being moved above the sensor. In contrast forward reach of arms can typically encompass 40-50~cm \cite{sanders1993applied}, which is a more common scenario for the Leap Motion controller mounted on a HMD.  Brown et al. \cite{brown2014performance} compared the performance of the Leap Motion to mouse in a Fitts'  Law task (with two confirmation methods) and found the throughput of the Leap Motion to be significantly lower than that of the mouse. Again, a table mounted setup was used for the Leap Motion. Tung et al. \cite{tung2015evaluation} indicated a mean pointing accuracy of the Leap Motion controller of 17.3 mm ($sd = 9.6$) in a study with human participants ($n=9$). Participants were pointing at 15 targets at a vertical monitor at an approximate distance of 35 cm. Again the leap motion was mounted flat on the supporting desk.  Valentini et al. \cite{valentini2017accuracy} employed a similar test setup with the Leap Motion lying on a table and ten users touching a glass plate mounted above the device at distances between 20 and 60~cm. They report mean tracking errors between 4 and 7 mm. Marin et al. proposed to combine the data from an XBox Kinect 1 sensor and the Leap Motion (again in a table mounted setup) \cite{marin2016hand} but did not report positional accuracy. Lindsey \cite{lindsey2017evaluation} compared a HMD-mounted Leap Motion Controller with other commodity input devices (gamepad, touchpad) in a human participants study (n = 23) but did not report accuracy measurements. Lindsey focused on time and errors as objective measures in a shape, color and texture matching task) and found the Leap Motion controller to be significantly slower to the other input devices and also to be the least preferred. Xiao et al. \cite{xiao2018mrtouch} evaluated a novel finger tracking algorithm for the Microsoft HoloLens2, but did not compare it against other devices. In a user study (n=17) they found the mean spatial accuracy of the system to be 5.4 mm (sd=3.2) with a systematic offset to the right of predefined touch targets. Han et al. \cite{han2020megatrack} developed and evaluated a hand tracking algorithm which can be assumed to be a basis for the current Oculus Rift hand tracking system. They reported a mean positional error of 11mm relative to a ground truth optical outside-in tracking system for a number for in-air hand movements. 

Complementary to these prior works, our study allows for direct comparison of the accuracy of multiple hand tracking systems specific to AR and VR HMDs in a joint experimental design.

Recently, Batmaz et al. \cite{batmaz2020touch} utilized a HMD-mounted Leap Motion to investigate human performance between a VR, AR and 2D touchscreen condition (with and without haptic feedback) in a eye-hand coordination reaction test. In a human participants study (n=15) they found the throughput of both the VR and AR condition to be significantly lower than in the 2d touchscreen condition and the throughput of AR to be also significantly lower compared to VR and hypothesized that the difference between AR and VR might be due to the display system (as the same hand tracking technology was used). The closest work to ours is by Schneider et al. \cite{schneider2020accuracy}, who compared the finger tracking accuracy of the HTC Vive Pro and the leap motion. Specifically, our work complements this prior work by including a larger set of both recent AR and VR displays as well as a more comprehensive evaluation using additional dependent variables such as workload, subjective feedback and usability ratings. Our work also includes a complementing robot study.




\section{Study with Human Participants}

In a study with human participants ($n=20$), we compared the overall spatial accuracy of different head-mounted displays. Specifically, we compared three solutions for hand tracking in VR HMDs and two AR systems.






\subsection{Experimental Design}
\label{sec:experimentalDesign}
\begin{figure}[t]
	\centering 
	\includegraphics[width=0.7\columnwidth]{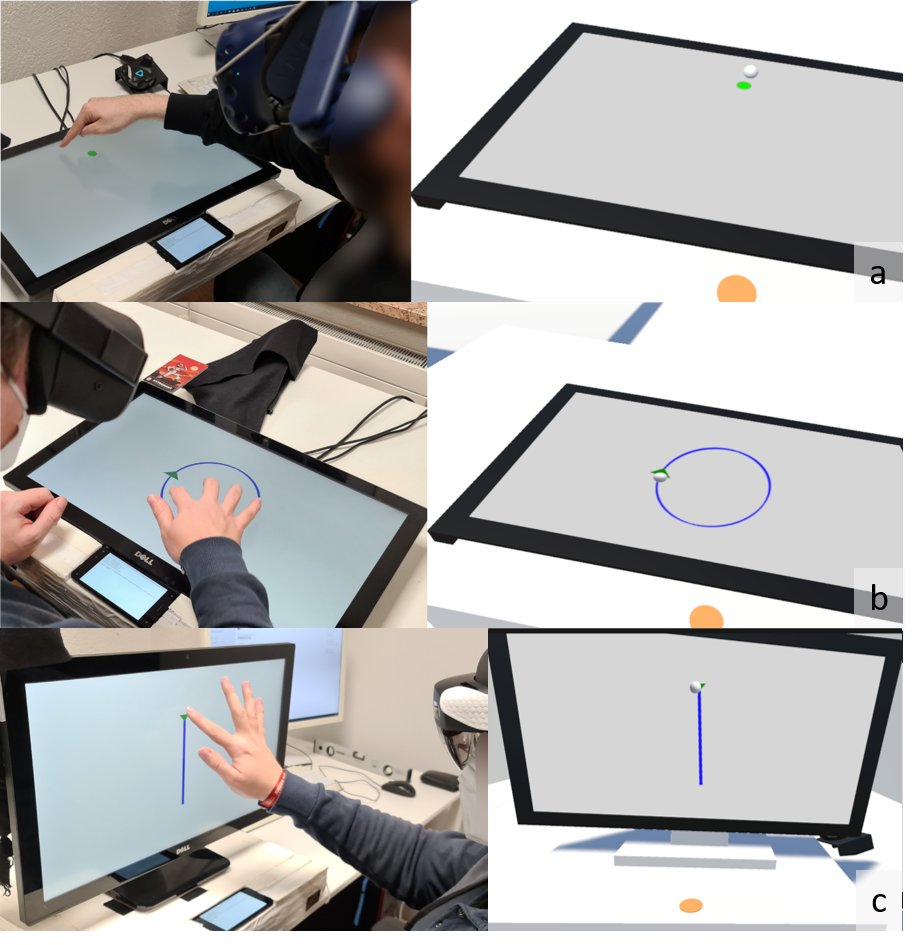}
	\caption{\small{Participants performing tasks (left) with respective VR view (right): a: \textsc{Target Acquisition} task in \textsc{Horizontal} with \textsc{LeapMotion}; b: \textsc{Shape Tracing} task in \textsc{Horizontal} with \textsc{Quest}; c: \textsc{Shape Tracing} in \textsc{Vertical} with \textsc{HoloLens}.}}
	\label{fig:allTasksExample}

\end{figure}


We conducted the an experiment \hl{consisting of two parts, one for VR devices and one for AR devices}.
The part comparing the VR systems was designed as a $3 \times 2$ within-subjects design.
The part comparing AR systems used a $2 \times 2$ within-subjects design.
The two independent variables in both parts were \textsc{Interface} and \textsc{Orientation}. 
\textsc{Interface} represented tested platforms. For the VR experiment these were: \textsc{Vive: V}, \textsc{Quest: Q} and \textsc{LeapMotion: LM}
For the AR experiment these were: \textsc{HoloLens: HL} and \textsc{MagicLeap: ML}. 
Each represents a commodity device or sensor listed in Section \ref{sec:apparatus}. 
Following prior work \cite{xiao2018mrtouch}), each participant performed the tasks in two different \textsc{Orientation}s:  \textsc{Horizontal} and \textsc{Vertical}. 
These two values refer to the orientation of the touchscreen, on which users had to perform the tasks and corresponds to typical interaction scenarios (e.g., wall interaction, desk interaction). \textsc{Horizontal} describes a flat configuration, where the screen was parallel to the table (see Figure \ref{fig:allTasksExample}, a) and \textsc{Vertical} refers to a standing configuration, where the screen is perpendicular to the table (see Figure \ref{fig:allTasksExample}, c). 
To mitigate ordering effects, the sequence of \textsc{Interface}s and \textsc{orientation} were counterbalanced between participants.

Dependent variables for both experiments (VR and AR) included workload, as measured by NASA TLX \cite{hart1988development}, Usability as measured by the System Usability Scale (SUS) \cite{brooke1996sus}) and Simulator Sickness as measured by the Simulator Sickness Questionnaire (SSQ) \cite{kennedy1993simulator}.

Based on prior work \cite{grubert2018effects}, we provided a three-item questionnaire (which we name Perceived Finger Assessment (PFA) within this paper) with a 7 point Likert scale (ranging from "totally disagree" to "totally agree"). The items were: "I felt that the fingers were my own." (PFA-F), "I felt that I could control the position of my fingers." (PFA-C) and "I felt that I hit the target I aimed for." (PFA-A).

\begin{figure}[t]
	\centering 
	\includegraphics[width=1\columnwidth]{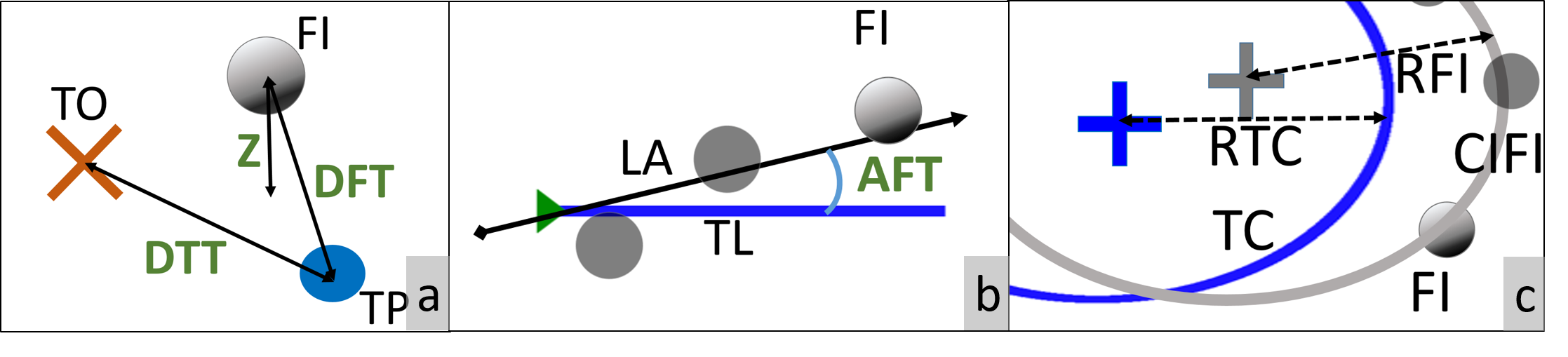}
	\caption{\small{Measures used for assessing spatial accuracy (highlighted in green). a: Metrics used in the \textsc{Target Acquisition} task with \textsc{Z}: \textsc{Z-Distance} between the tracked finger (FI) and the display surface, \textsc{DFT}: \textsc{Distance Finger-Target} between FI and the target point (TP), \textsc{DTT}: \textsc{Distance Touch-Target} between the touch position (TO) and TP;
	b and c: Metrics used in the \textsc{Shape Tracing Task} with \textsc{AFT}: \textsc{Angle Finger-Target} being the angle between the line fitted through the tracked finger points (LA) and the target line (TL), \textsc{Radius Difference Finger-Target} is the difference between the radius (RFI) of the circle (CIFI) fitted through the FIs and the radius (RTC) of the target circle (TC). }
	}
	
	
	\label{fig:explanation}
\end{figure}

Furthermore, the spatial accuracy was assessed using the following measures, see also Figure \ref{fig:explanation}: 
\textsc{Distance Finger-Target} describes the Euclidean distance between the index finger position as tracked by the system and the target on the screen while touching \hl{and is an indicator for the spatial accuracy of the tracking system}. 
\textsc{Distance Touch-Target} describes the distance between the touch position on the screen and the target while touching. \hl{This metric was chosen as it describes the offset between the actual touch position and the position recognized by the target position.}
\textsc{Z-Distance} is the distance between the finger as tracked by the respective system and the plane representing the screen while the participant is touching.
\textsc{Angle Finger-Target} describes the angle between the displayed target line and a line that is  fitted through the tracked finger positions while the user was tracing the line.
\textsc{Radius Difference Finger-Target} is the difference between the radius of the target circle and the circle that was fitted through the tracked finger positions while the user was tracing the circle.
\hl{\textsc{Angle Finger-Target} and \textsc{Radius Difference Finger-Target} were chosen to assess the relative accuracy of tracking a moving finger. Other possible measures like circle center or start and endpoint of the target line were not considered because the absolute accuracy was already assessed through \textsc{Distance Finger-Target}, \textsc{Distance Touch-Target} and \textsc{Z-Distance}.}



\subsection{Tasks}
\label{sec:task}

\begin{figure}[t]
	\centering 
	\includegraphics[width=1\columnwidth]{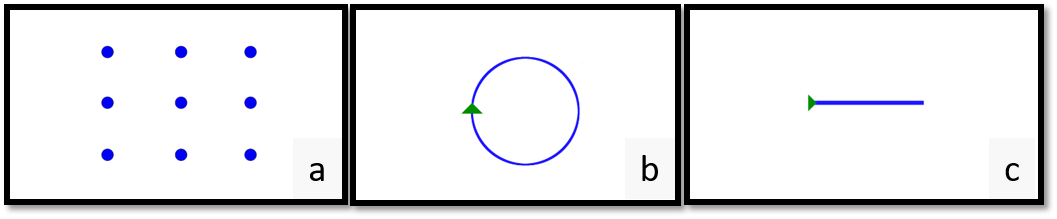}
	\caption{\small{a: \textsc{Target Acquisition} task with 9 possible target locations displayed simultaneously for visualization. \textsc{Shape Tracing} task displays two shapes: b: a circle to be traced clockwise and c: a line to be traced from left to right.}}
	\label{fig:TasksOnePic}

\end{figure}



The participants were told to conduct two tasks in all conditions, inspired by prior work \cite{harrison2011omnitouch, xiao2018mrtouch, schneider2020accuracy}. 
For each task, the participants started with their index finger on the smartphone in front of them, see Figure \ref{fig:allTasksExample}, left. The whole smartphone display served as home position. The home position in VR was represented as orange dot within the bounds of the smartphone (the smartphone was not visualized in VR), see Figure \ref{fig:allTasksExample}, right, i.e. participants could still touch slightly to the left or right of the visualized homing position in VR.

In the \textsc{Target Acquisition} task, participants needed to touch a blue round target (diameter: 18 mm) on a touchscreen in front of them as soon as it appeared and hold it with the tip of their dominant hand's index finger for about one second until its color changed from blue to green. 
The request for holding the fingertip for a short period enables the collection of multiple tracking samples at the target position. After touching the target, participants had to return their index finger to the resting position on the smartphone.  The next target was shown as soon as the finger touched the smartphone surface. The return to the smartphone had two reasons. First, the touch on the smartphone was used to trigger the visualization of the next target, and, second, to ensure the movement of the hand for each position is similar. 
When using the \textsc{VR-HMD}s, the participants could not see the real world including the target touchscreen, the phone or even their own hands. Therefore, the tip of their index finger was rendered as a sphere with a diameter of 16~mm (following recommendation by prior work \cite{grubert2018effects}). Using a sphere with 16mm allowed to position it completely within the target circle having a diameter of 18~mm. 
Additionally, the participants saw 3D models of the table and the touchscreen. 
When using \textsc{AR-HMD}s there was no virtual rendering of the participant's hands or any other objects to avoid potential confusion between real and tracked fingertips when touching a target.

One trial in the \textsc{Target Acquisition} task consisted of hitting a total number of nine circular targets, which appeared in a fixed order left to right and top to bottom (see Figure \ref{fig:TasksOnePic}, a). \hl{As we focused on pointing accuracy and not target acquisition time, randomizing the order of targets was not necessary.}
The distance between the center target and the starting position on the phone is 25.9 cm for the \textsc{Vertical} orientation and 20.1 cm for \textsc{Horizontal}. The other eight targets have an offset from the center target of 7.5 cm to the left (or right) and 7.5 cm to the top or bottom. 
Participants were asked to conduct five trials, with nine targets each, for a total of 5 trials $\times$ 9 targets $\times$ 3 interfaces 2 $\times$ orientations = 270 recorded target acquisitions per user for the VR experiment and 5 trials $\times$ 9 targets $\times$ 2 interfaces 2 $\times$ orientations = 180 recorded target acquisitions per user for the AR experiment.

The second task, \textsc{Shape Tracing}, also starts with placing the tip of the index finger on the phone's touchscreen. 
On the target touchscreen a blue shape was displayed, being either a circle with a radius of 7.5 cm or a straight 15 cm long line. A green triangle on the shape indicated the point that the users should touch with their index fingertip and follow the shape in the direction displayed by the triangle (see Figure \ref{fig:TasksOnePic}, b and c). The circles 
had to be traced both clockwise and counterclockwise. The line segments had to be traced from left-to-right, right-to-left, top-to-bottom and bottom-to-top. Again, after the participants finished tracing one shape, they were asked to return their index finger to the phone screen. 
The \textsc{Shape Tracing} task also consisted of five trials. One trial consisted of two circles (clockwise and counterclockwise) and four lines (in all four directions).  
\hl{Again, the order of the target shapes was not randomized, as the experiment focused on accuracy rather than speed.}

\subsection{Procedure and Data Collection}
After an introduction, participants were asked to fill out a demographic questionnaire. The study was divided into two parts. In one part, \textsc{VR-HMDs} were tested, and in another the \textsc{AR-HMDs}. The order of the parts as well as the order of the devices inside the parts were counter-balanced, as well as the order of \textsc{Orientation}. 
Before starting with the tasks in a specific \textsc{orientation}, a calibration step mapped the position and orientation of the touchscreen to the respective HMD coordinate system. Depending on the \textsc{Interface}, different techniques were used for the calibration process:  \textsc{LeapMotion} and \textsc{Vive} where calibrated using a Vive Tracker that could be held to the corner of the screen and its position in 6DOF to the Vive coordinate system. \textsc{Quest} was calibrated with a modified Oculus controller (for pictures of the calibration tools see the appendix). 
Following the calibration process, participants were asked to conduct a training session before starting the data collection session. With each touch on the touchscreen the touch data are sent via WiFi to the HMD. The HMD converted the 2D touch position of the monitor to 3D touch positions in the virtual world and saved with the target position and the index finger position in a file. At the beginning of each stage (being either \textsc{Horizontal} or \textsc{Vertical} \textsc{Orientation}) the participants were asked to sit in a comfortable position. 
After conducting all tasks (\textsc{Target Acquisition} and \textsc{Shape Tracing}) in both \textsc{Horizontal} and \textsc{Vertical} orientation, the participants filled out the SSQ, SUS and TLX questionnaires 
as well as the three questions for perceived finger assessment once for each HMD. Following the completion of all three \textsc{VR-HMD}s, the participants were asked which HMD they preferred.  
This was not done after AR conditions, because participants saw their real hand and the real touchscreen (no augmentations). Therefore, they were unlikely to have a preference related to the tracking performance. The duration of the study was about 110 minutes per participant. They were compensated with a voucher worth \texteuro10.   

For the target acquisition task, we collected multiple touch samples (at minimum 20 samples, more if the users touched down longer) per target location, which were averaged into one. For the shape tracing task, we sampled touch (and the associated finger) positions at a distance of at least .01~mm to the previous sample. In this way, we ensure that the start (and end) positions of the drawn shapes only consist of single points, even if the participant's finger rests for a little while on that position.  This is important for the fitting of the line and circle to not bias the start and end. For the line fitting we used \hl{linear optimization and} the RANSAC algorithm \cite{fischler1981random} for robust line fitting. For fitting the circle, we utilized the singular value decomposition and the method of least-squares for the optimal circle fitting.  

\subsection{Apparatus}
\label{sec:apparatus}



The following HMDs were employed: \textsc{{\bf HoloLens}}: A Microsoft HoloLens 2 AR-HMD (see Figure \ref{fig:teaser} a. bottom-left) running Windows 19041.1136. \textsc{{\bf MagicLeap}}: A Magic Leap 1 AR-HMD (see Figure \ref{fig:teaser} a. bottom-right) running LuminOS 0.98.20. \textsc{{\bf Quest}}: An Oculus Quest 1 VR-HMD (see Figure \ref{fig:teaser} a. top-center) running Quest OS 25.0.0.77. \textsc{{\bf Vive}}: An HTC VIVE Pro VR-HMD (see Figure \ref{fig:teaser} a. top-left) using VIVE Hand Tracking SDK 0.9.4. \textsc{{\bf LeapMotion}}: An Ultraleap LeapMotion sensor attached to a VIVE Pro VR-HMD  (see Figure \ref{fig:teaser} a. top-right). It uses LeapMotion Software 4.1.0. The VIVE Pro HMD is similar to the one used in \textsc{Vive}, without using its built-in hand tracking capabilities.

Both AR-HMDs used the image marker tracking provided by Vuforia 9.7.5 using a default image, to allow a better comparison between the devices. Also both HTC VIVE Pro devices were used with Steam VR 1.16.8 and Steam VR Unity Plugin 2.7.3 (SDK version 1.14.15). 
The system for all VR \textsc{Interfaces} was implemented in Unity 2020.2 and deployed on a PC (Intel Xeon E5-2620 v 4 processor, 64 GB RAM, Nvidia GTX 1060 graphics card) running Windows 10. The touchscreen was a Dell S2340T monitor with a screen width of 56.2 cm and a height of 34 cm. The smartphone for the resting position was a Fire Phone with a screen width of 10.4 cm and a height of 5.8 cm.

Participants were seated on a standard office chair (seat height between a minimum of 46.5 cm and a maximum of 57 cm), adjusted to a comfortable height for each user. They were asked to initially place their index fingertip of their dominant hand on a touchscreen of a phone that lay on the table in front of them (See Figure \ref{fig:allTasksExample}).
For the communication between touchscreen, smartphone and the HMD-device, we implemented an application to send the touch information to the respective device via WiFi and UDP protocol. 

\subsection{Participants}
We recruited 20 participants (11 female, 9 male, mean age 33.4 years, sd = 8.9) with diverse backgrounds. All participants were familiar with touch sensitive screens. Skin types (which might influence tracking capabilities) were ranging from I to IV according to Fitzpatrick \cite{fitzpatrick1975soleil}. Three of the participants indicated no prior VR experience, 11 participants rarely used VR, yet more than once, two participants occasionally, two often and two participant very frequently. Four participants indicated no prior AR experience, 11 had rarely used AR yet more than once, two participants occasionally, one often and two participant very frequently. 
Eight participants wore contact lenses or glasses which corrected to normal vision. 19 participants were right handed while one was left handed. All participants used the index finger of their dominant hand to conduct the tasks.


\subsection{Results}
Unless otherwise specified, statistical significance tests for performance data (task completion time, distance to target) were carried out using general linear model repeated measures analysis of variance (RM-ANOVA) with Holm-Bonferroni adjustments for multiple comparisons at an initial significance level $\alpha = 0.05$. We indicate effect sizes whenever feasible ($\eta^2_p$). For subjective feedback, or for data that did not follow a normal distribution or could not be transformed to a normal distribution using a log-transform, we used Friedman's test with Holm-Bonferroni adjustments for multiple comparisons using Wilcoxon signed-rank tests. The anonymized raw and aggregated data of the study are available under \newline https://gitlab.com/mixedrealitylab/finger-tracking-accuracy. 

Selected additional results are depicted in the appendix.


\subsubsection{AR}

\begin{figure}[t]
	\centering 
	\includegraphics[width=\columnwidth]{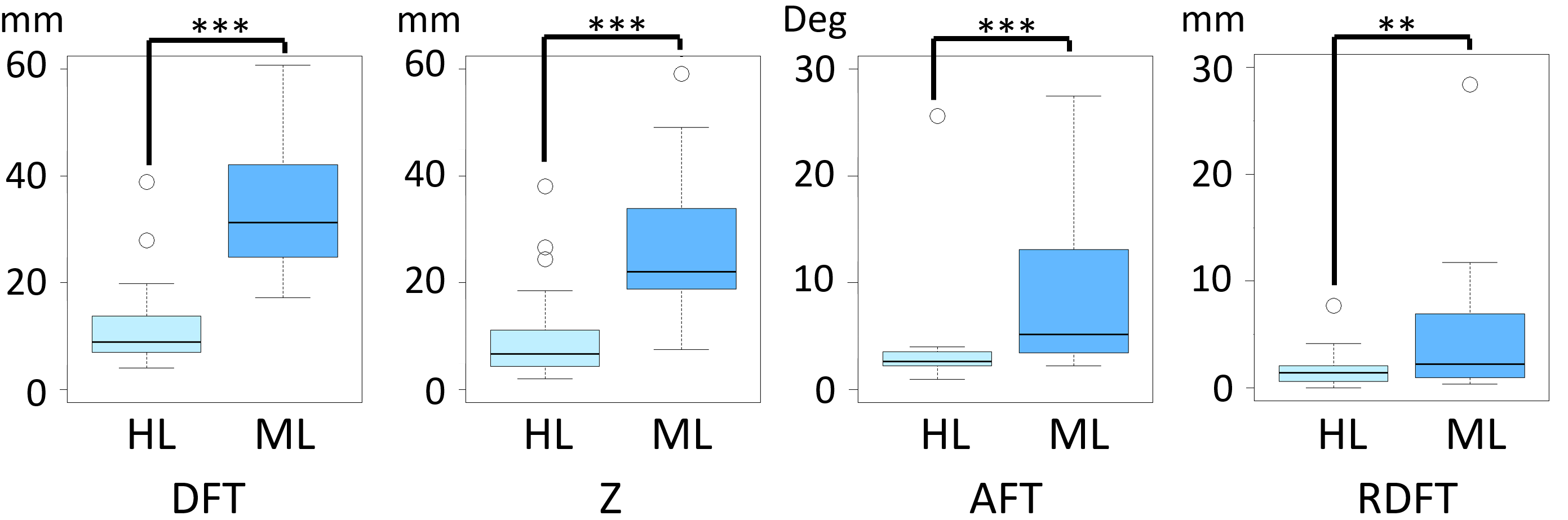}
	\caption{\small{Accuracy measures from the human participants study for the \textsc{HoloLens} (HL) and \textsc{MagicLeap} (ML); DFT = Distance Finger-Target in millimeter; Z = Z-Distance in millimeter; AFT = Angle Finger-Target in degrees; RDFT = Radius Difference Finger-Target in millimeter; * = $p<0.05$; ** = $p<0.01$; *** = $p<0.001$.}}
	\label{fig:UserStudy-AR-Boxplot}

\end{figure}

For the \textsc{Target Acquisition Task} we analyzed the distance between the tracked finger position and the displayed target (\textsc{Distance Finger-Target: DFT}) and the distance between the tracked finger position and the display-area (\textsc{Z-Distance: Z}).
We did not analyze the distance between the touch position on the display and the displayed target (\textsc{Distance Touch-Target: DTT}), because this measure does not tell us anything about the tracking accuracy of the AR-devices (the user aligned their real fingertip with the target, not the virtual fingertip as in the VR HMDs). 
The descriptive statistics and the results from the null hypothesis significance tests (NHST) are presented in Table \ref{tab:performance_ratings_AR} and Figure \ref{fig:UserStudy-AR-Boxplot}. \hl{For the significance tests, the data was log-transformed to ensure a normal-distribution.}
The results indicate that the \textsc{Interface} significantly influenced both the \textsc{Distance Finger-Target} and the \textsc{Z-Distance}, in such a way, that the \textsc{HoloLens} was more accurate than the \textsc{MagicLeap}.
The results show no significant influence of \textsc{Orientation} on the accuracy.

\begin{table*}[t]
    \centering 
    \tiny
    \begingroup

        \begin{tabular}{|>{\centering}m{0.45cm}||>{\centering}m{0.2cm}|>{\centering}m{0.2cm}|c|c|c||>{\centering}m{0.2cm}|>{\centering}m{0.2cm}|c|c|c||>{\centering}m{0.2cm}|>{\centering}m{0.2cm}|c|c|c||>{\centering}m{0.2cm}|>{\centering}m{0.2cm}|c|c|c|} 
            \multicolumn{1}{c}{}& \multicolumn{10}{c}{\small\bfseries\textbf{AR - Target Acquisition Task }} & \multicolumn{10}{c}{\small\bfseries\textbf{AR - Shape Tracing Task}} \\
            \hline 
            
            & \multicolumn{5}{c||}{Distance Finger-Target} &\multicolumn{5}{|c||}{Z-Distance} & \multicolumn{5}{c||}{Angle Finger-Target} &\multicolumn{5}{|c|}{Radius Difference Finger-Target}\\
            \hline 
            &  d$f_{1}$ & d$f_{2}$ & F & p &  $\eta^2_p$ & d$f_{1}$ & d$f_{2}$ & F & p &  $\eta^2_p$ & d$f_{1}$ & d$f_{2}$ & F & p &  $\eta^2_p$ & d$f_{1}$ & d$f_{2}$ & F & p &  $\eta^2_p$  \\ 
            \hline 

            I &   \cellcolor{lightgray}$1$ &  \cellcolor{lightgray}$19$ &  \cellcolor{lightgray}$91.3542$ & \cellcolor{lightgray}$<.001$ &  \cellcolor{lightgray}$.828$  &  \cellcolor{lightgray}$1$ &  \cellcolor{lightgray}$19$ &  \cellcolor{lightgray}$64.612$ &  \cellcolor{lightgray}$<.001$ &  \cellcolor{lightgray}$.764$ & \cellcolor{lightgray} $1$ & \cellcolor{lightgray} $19$ & \cellcolor{lightgray} $23.46$ & \cellcolor{lightgray} $<.001$ & \cellcolor{lightgray} $.55$ & \cellcolor{lightgray} $1$ & \cellcolor{lightgray} $19$ & \cellcolor{lightgray} $11.846$ & \cellcolor{lightgray} $.003$ & \cellcolor{lightgray} $.384$  \\ 
            
            \hline
            
            O &   $1$ & $19$ & $.0281$ & $.869$ & $.001$  &  $1$ &  $19$ &  $.134$ &  $.719$ &  $.007$  & $1$ & $19$ & $.0$ & $.975$ & $0.00$  &  $1$ &  $19$ & $.0483$ &  $.82$ &  $.003$ \\ 

            \hline 

            \tiny{I $\times$ O} &   $1$ & $19$   & $1.4861$ & $.238$  & $.073$ &  $1$ &  $19$ & $.357$ &  $.557$ & $.018$  & $1$ & $19$   & $4.20$ & $.054$  & $.18$ & \cellcolor{lightgray} $1$ & \cellcolor{lightgray}  $19$ & \cellcolor{lightgray} $11.126$ &\cellcolor{lightgray}  $.003$ & \cellcolor{lightgray} $.363$  \\ 

            \hline 

        \end{tabular}
    
        \begin{tabular}{|c|c|c|c|c|c|c|c|c|} 
             \multicolumn{1}{c}{} & \multicolumn{4}{c}{\small\bfseries\textbf{AR - Target Acquisition Task }} & \multicolumn{4}{c}{\small\bfseries\textbf{AR - Shape Tracing Task}} \\
            \hline 
            & \multicolumn{2}{c|}{Distance Finger-Target} & \multicolumn{2}{c|}{Z-Distance} & \multicolumn{2}{c|}{Angle Finger-Target} & \multicolumn{2}{c|}{Radius Difference Finger-Target}  \\
            \hline 
            Device & Mean & SD & Mean & SD & Mean & SD & Mean & SD  \\ 
            \hline 

            \textsc{HoloLens} & $14.54~mm$ & $9.22$ &  $11.8~mm$ & $19.15$ & $3.29$° & $1.31$ & $1.64$~mm  & $1.47$  \\ 
            
            \hline
            
            \textsc{MagicLeap} & $39.84~mm$ & $16.11$ & $29.85~mm$  & $19.99$ & $9.55$° & $7.4$ & $8.62$~mm  & $20.39$  \\ 

            \hline 

        \end{tabular}

    \endgroup

    \caption{RM-ANOVA results for the human participants study and mean and standard-deviation values for AR-devices (regardless of \textsc{Orientation}).  Gray rows show significant findings. I = \textsc{Interface}, O = \textsc{Orientation}. }
    \label{tab:performance_ratings_AR}
\end{table*}


The \textsc{Shape Tracing Task} consisted of two types of tasks, tracing a line and tracing a circle. For the line tracing we analyzed the \textsc{Angle Finger-Target: AFT} which describes the angle between the target line and the line fitted through the tracked finger positions.  For the circle tracing we analyzed the \textsc{Radius Difference Finger-Target}, which describes the difference between the radius of the target circle and the radius of the circle fitted through the tracked finger points. These metrics are depicted in Figure \ref{fig:explanation}, b and c. The descriptive statistics and the NHST results are presented in Table \ref{tab:performance_ratings_AR}. \hl{For the significance tests, the data was log-transformed to ensure a normal-distribution.} For both metrics the results indicate a significant influence of \textsc{interface} on the measures. Post-hoc tests indicate that for both metrics the \textsc{HoloLens} is significantly more accurate than the \textsc{MagicLeap}, which means that the direction of the fitted line and the radius of the fitted circle are more similar to the target shapes. For \textsc{Radius Difference Finger-Target} the analysis also indicates interaction effects, because the \textsc{HoloLens} performs slightly better in \textsc{Horizontal} orientation and \textsc{MagicLeap} slightly better in \textsc{Vertical} orientation. However, post-hoc test did not reveal significant differences. \textsc{Workload and Usability:} The results of the four questionnaires showed no significant differences between the two AR-HMDs regarding usability, simulator sickness, task load or perceived finger assessment (see also appendix).

\subsubsection{VR}

\begin{figure}[t]
	\centering 
	\includegraphics[width=1.0\columnwidth]{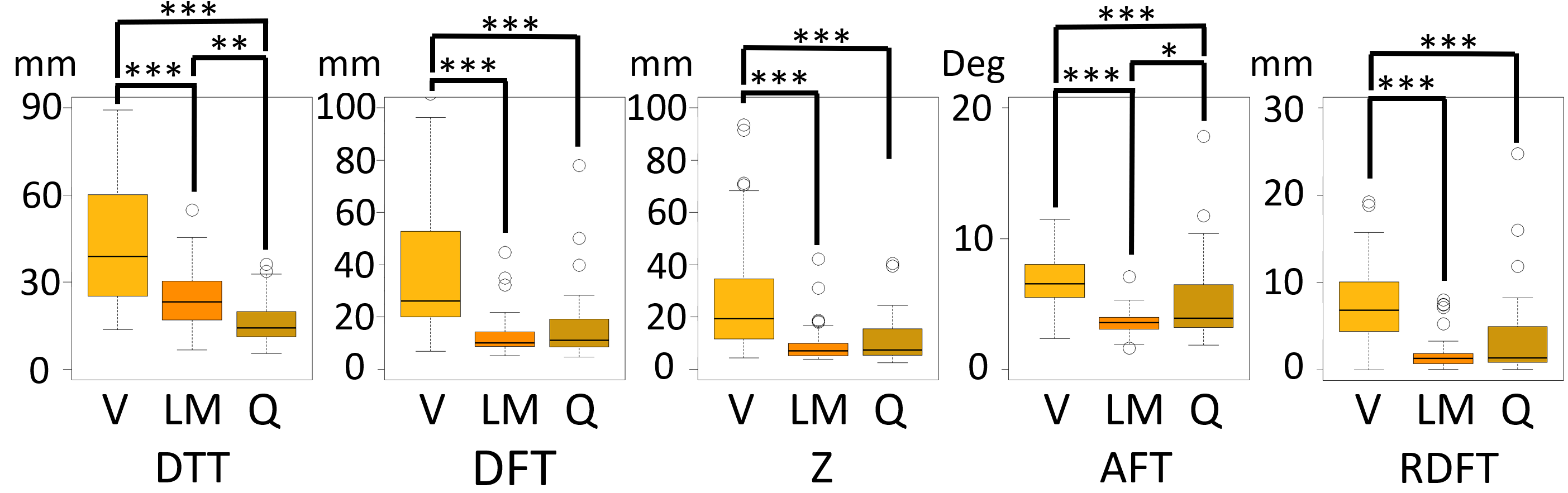}
	\caption{\small{Accuracy measures from human participants study for the \textsc{Vive} (V), the \textsc{LeapMotion} (LM) and \textsc{Quest} (Q). DTT = Distance Touch-Target in millimeter; DFT = Distance Finger-Target in millimeter; Z = Z-Distance in millimeter; AFT = Angle Finger-Target in degrees; RDFT = Radius Difference Finger-Target in millimeter; * = $p<0.05$; ** = $p<0.01$; *** = $p<0.001$.}}
	\label{fig:UserStudy-VR-Boxplot}

\end{figure}

For the \textsc{Target Acquisition Task} we analyzed the distance between the tracked finger position and the displayed target (\textsc{Distance Finger-Target}), between the touch position on the display and the displayed target (\textsc{Distance Touch-Target}) and the distance between the tracked finger position and the display-area (\textsc{Z-Distance}).
These metrics are depicted in Figure 6. \hl{For the significance tests, the data was log-transformed to ensure a normal-distribution.}
The descriptive statistics and the NHST results are presented in Table \ref{tab:performance_ratings_VR} and Figure \ref{fig:UserStudy-VR-Boxplot}.
The results indicate that the \textsc{Interface} has a significant influence on \textsc{Distance Touch-Target} and \textsc{Distance Finger-Target}.
Post-hoc tests show that for both metrics \textsc{Vive} is significantly less accurate than both \textsc{Quest} and \textsc{LeapMotion}.
In addition, \textsc{LeapMotion} is significantly less accurate than \textsc{Quest} for the \textsc{Distance Touch-Target} measure.
For both measures an interaction effect is detected. The results indicate that the \textsc{LeapMotion} and \textsc{Quest} are more accurate in the \textsc{Horizontal} orientation, while \textsc{Vive} is more accurate in the \textsc{Vertical} orientation.
Analyzing the \textsc{Z-Distance} results indicates a significant influence of \textsc{Interface} and \textsc{Orientation} as well as interaction effects. Post-hoc tests indicate that, again, \textsc{Vive} is less accurate than \textsc{LeapMotion} and \textsc{Quest}.
They also indicate that the \textsc{horizontal} orientation ($M=10.87$~mm, $SD=8.07$) is more accurate than the \textsc{Vertical} orientation ($M=21.38$~mm, $SD=22.08$).

\begin{table*}[t]
    \centering 
   \tiny
    \begingroup
    \setlength{\tabcolsep}{5pt}

     \begin{tabular}{|>{\centering}m{0.45cm}||>{\centering}m{0.15cm}|>{\centering}m{0.15cm}|>{\centering}m{0.4cm}|>{\centering}m{0.5cm}|>{\centering}m{0.2cm}||>{\centering}m{0.15cm}|>{\centering}m{0.15cm}|>{\centering}m{0.4cm}|>{\centering}m{0.5cm}|>{\centering}m{0.2cm}||>{\centering}m{0.15cm}|>{\centering}m{0.15cm}|>{\centering}m{0.4cm}|>{\centering}m{0.5cm}|>{\centering}m{0.2cm}||>{\centering}m{0.15cm}|>{\centering}m{0.15cm}|>{\centering}m{0.4cm}|>{\centering}m{0.5cm}|>{\centering}m{0.2cm}||>{\centering}m{0.15cm}|>{\centering}m{0.15cm}|>{\centering}m{0.4cm}|>{\centering}m{0.5cm}|c|}  
     
            \multicolumn{1}{c} {} & \multicolumn{15}{c}{\small\bfseries\textbf{VR - Target Acquisition Task }}  &  \multicolumn{10}{c}{\small\bfseries\textbf{VR - Shape Tracing Task }} \\
            \hline 
            & \multicolumn{5}{c||}{Distance Touch-Target} &\multicolumn{5}{|c||}{Distance Finger-Target} & \multicolumn{5}{|c||}{Z-Distance} & \multicolumn{5}{c||}{Angle Finger-Target} &\multicolumn{5}{|c|}{Radius Diff. Finger-Target} \\
            \hline 
            &  d$f_{1}$ & d$f_{2}$ & F & p &  $\eta^2_p$ & d$f_{1}$ & d$f_{2}$ & F & p &  $\eta^2_p$ & d$f_{1}$ & d$f_{2}$ & F & p &  $\eta^2_p$ & d$f_{1}$ & d$f_{2}$ & F & p &  $\eta^2_p$ & d$f_{1}$ & d$f_{2}$ & F & p &  $\eta^2_p$     \\ 
            \hline 

            I &  \cellcolor{lightgray} $2$ &  \cellcolor{lightgray}$38$ & \cellcolor{lightgray} $42.81$ & \cellcolor{lightgray}$<.001$ &  \cellcolor{lightgray}$.69$  & \cellcolor{lightgray} $2$ &  \cellcolor{lightgray}$38$ &  \cellcolor{lightgray}$50.8$ &  \cellcolor{lightgray}$<.001$ &  \cellcolor{lightgray}$.73$ & \cellcolor{lightgray} $2$ & \cellcolor{lightgray} $38$ & \cellcolor{lightgray} $25.83$ & \cellcolor{lightgray} $<.001$ & \cellcolor{lightgray} $.58$ & \cellcolor{lightgray} $2$ & \cellcolor{lightgray} $38$ & \cellcolor{lightgray} $27.73$ & \cellcolor{lightgray} $<.001$ & \cellcolor{lightgray} $.593$ & \cellcolor{lightgray} $2$ & \cellcolor{lightgray} $38$ & \cellcolor{lightgray} $21.411$ & \cellcolor{lightgray} $<.001$ & \cellcolor{lightgray} $0.53$   \\ 
            
            \hline
            
            O &  $1$ &  $19$ &  $1.76$ & $.2$ & $.09$  &  $1$ &  $19$ &  $.0$ &  $.99$ &  $.0$ & \cellcolor{lightgray} $1$ & \cellcolor{lightgray}$19$ & \cellcolor{lightgray} $9.11$ & \cellcolor{lightgray} $.007$ & \cellcolor{lightgray} $.32$  & $1$ & $19$ & $0.002$ & $0.96$ & $0$  &  \cellcolor{lightgray}$1$ &  \cellcolor{lightgray}$19$ & \cellcolor{lightgray}$5.406$ &  \cellcolor{lightgray}$0.031$ &  \cellcolor{lightgray}$0.222$  \\ 

            \hline 

            \tiny{I $\times$ O} &  \cellcolor{lightgray}$2$ & \cellcolor{lightgray} $38$ & \cellcolor{lightgray} $25.13$ & \cellcolor{lightgray} $<.001$ & \cellcolor{lightgray} $.57$  &  \cellcolor{lightgray}$2$ & \cellcolor{lightgray} $38$ & \cellcolor{lightgray} $11.8$ &  \cellcolor{lightgray}$<.001$ & \cellcolor{lightgray} $.38$ & \cellcolor{lightgray} $1$ & \cellcolor{lightgray} $38$ & \cellcolor{lightgray} $10.24$ & \cellcolor{lightgray} $<.001$ & \cellcolor{lightgray} $.38$ & \cellcolor{lightgray}  $2$ & \cellcolor{lightgray} $38$   & \cellcolor{lightgray} $10.83$ & \cellcolor{lightgray} $<.001$  & \cellcolor{lightgray} $.36$ & $2$ & $38$ & $0.781$ & $0.465$ & $0.039$   \\ 

            \hline 

        \end{tabular}
        
        \begin{tabular}{|c|c|c|c|c|c|c|c|c|c|c|} 
            \multicolumn{1}{c} {} & \multicolumn{6}{c}{\small\bfseries\textbf{VR - Target Acquisition Task}}  &  \multicolumn{4}{c}{\small\bfseries\textbf{VR - Shape Tracing Task }} \\
            \hline 
            &  \multicolumn{2}{|c|}{Distance Touch-Target} & \multicolumn{2}{|c|}{Distance Finger-Target} & \multicolumn{2}{|c|}{Z-Distance} & \multicolumn{2}{c|}{Angle Finger-Target} & \multicolumn{2}{c|}{Radius Difference Finger-Target}  \\
            \hline 
            Device & Mean & SD & Mean & SD & Mean & SD & Mean & SD & Mean & SD    \\ 
            \hline 

            \textsc{vive} & $45.04$~mm & $19.51$ & $37.22$~mm & $20.73$ & $27.67$~mm & $23.98$ & $6.76$° & $1.67$ & $7.76$~mm  & $4.80$  \\ 
            
            \hline
            
           \textsc{LeapMotion} & $24.05$~mm & $8.13$ & $13.15$~mm & $5.45$  & $9.56$~mm & $7.57$  & $3.61$° & $0.65$ & $1.94$~mm  & $2.14$ \\ 

            \hline 

           \textsc{Quest} & $16.14$~mm & $5.81$ & $16.22$~mm & $8.76$ & $11.14$~mm & $9.1$ & $5.17$° & $2.44$ & $3.34~mm$ & $4.81$  \\ 

            \hline 

        \end{tabular}
    
    \endgroup

    \caption{\small{RM-ANOVA results for the human participants study and mean and standard-deviation values for VR-devices (regardless of \textsc{Orientation}).  Gray rows show significant findings. I = \textsc{Interface}, O = \textsc{Orientation}. }}
    \label{tab:performance_ratings_VR}

\end{table*}


For the \textsc{Shape Tracing Task}, the same measures as for the AR HMDs were employed. 
These metrics are depicted in Figure \ref{fig:explanation}, a. \hl{For the significance tests, the data was log-transformed to ensure a normal-distribution.} 
The mean values and results from analyzing these metrics are presented in Table \ref{tab:performance_ratings_VR}.
Analyzing the results indicates that \textsc{Interface} significantly influences the two metrics.
Post-hoc tests showed that for both the \textsc{Angle Finger-Target} and \textsc{Radius difference Finger-Target} the results from the \textsc{Vive} differed significantly more from the target shapes than \textsc{LeapMotion} and \textsc{Quest}. The analysis also indicated that \textsc{Orientation} significantly influences the \textsc{Radius difference Finger-Target}. Post-hoc tests indicate that the \textsc{Horizontal} orientation ($M=3.75$~mmm $SD=3.94$) matches the radius closer than the \textsc{Vertical} orientation ($M=4.94$~mmm $SD=5.47$). 
Additionally, regarding \textsc{Angle Finger-Target}, \textsc{LeapMotion} differed significantly less from the target shapes than \textsc{Quest}. Analyzing \textsc{Angle Finger-Target} also indicated interaction effects. Post-hoc tests showed that the \textsc{Quest} differed significantly less from the target line in the \textsc{Horizontal} compared to the \textsc{Vertical} condition. 

\begin{figure}[t]
	\centering 
	\includegraphics[width=0.65\columnwidth]{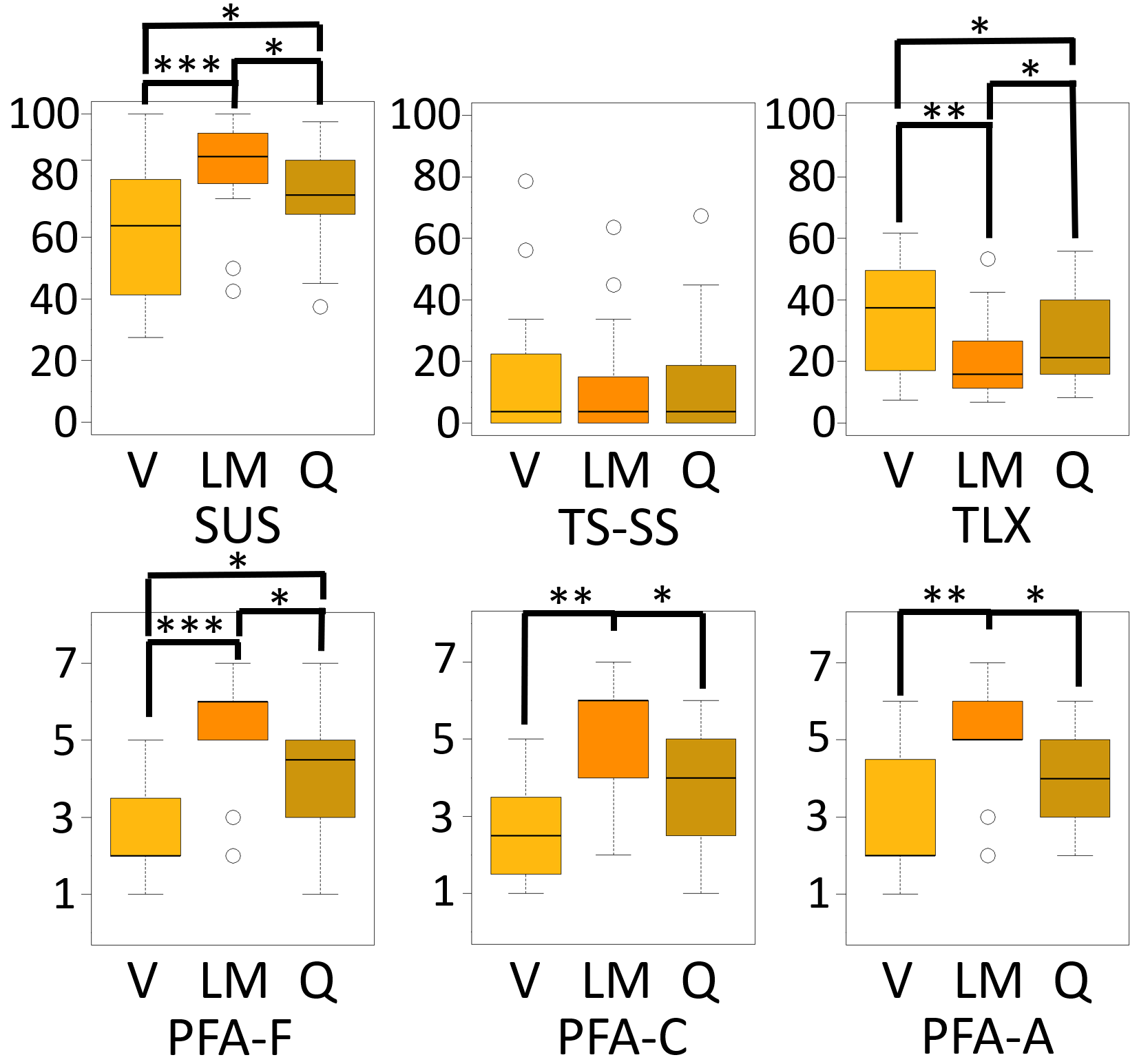}
	\caption{\small{Results from Questionnaires for \textsc{Vive} (V), \textsc{LeapMotion} (LM) and \textsc{Quest} (Q). SUS = System Usability Scale; TS-SS = Total Severity Simulator Sickness; TLX = Overall Taskload; PFA-F = Perceived Finger; PFA-C = Perceived Control; PFA-A = Perceived Accuracy; * = $p<0.05$; ** = $p<0.01$; *** = $p<0.001$.}
	\label{fig:UserStudy-VR-Boxplot-Questionnaire}}

\end{figure}

\textsc{Workload and Usability}: The analyzed questionnaire results for the three VR-HMDs and the mean and standard-deviation values are displayed in Table \ref{tab:questionnaires_VR} and Figure \ref{fig:UserStudy-VR-Boxplot-Questionnaire}. No significant influence of \textsc{Interface} could be found on the simulator sickness measurement. Regarding usability, \textsc{LeapMotion} received the highest scores, followed by \textsc{Quest} and then \textsc{Vive}. This aligns with the perceived overall taskload, where \textsc{Vive} resulted in the highest taskload, followed by \textsc{Quest} and the least taskload was recorded for \textsc{LeapMotion}. The results from the perceived finger assessment also indicate the highest scores for \textsc{LeapMotion}, second highest for \textsc{Quest} and least for \textsc{Vive}.

\begin{table*}[t]
    \centering 
    \small
    \begingroup
    \setlength{\tabcolsep}{5pt}

    \begin{tabular}{|c|c|c|c|c|c||c|c|c|c|c|c|}
        \multicolumn{12}{c}{\small\bfseries\textbf{Questionnaire Results for VR-HMDs}} \\

        \hline 
        & \multicolumn{5}{c||}{RM-Anova} & \multicolumn{2}{c|}{\textsc{vive}}  & \multicolumn{2}{c|}{\textsc{LeapMotion}}  & \multicolumn{2}{c|}{\textsc{Quest}}    \\ 
        
        \hline 
        & d$f_{1}$ & d$f_{2}$ & F & p &  $\eta^2_p$ & Mean & SD &  Mean & SD &  Mean &  SD   \\ 
        \hline 
        
        SUS &  \cellcolor{lightgray}$2$ & \cellcolor{lightgray}$38$ & \cellcolor{lightgray}$21.389$ & \cellcolor{lightgray}$<0.001$ & \cellcolor{lightgray}$0.53$ & $60.63$ & $22.08$ & $83.63$ & $15.36$ & $73.63$ &  $16.03$     \\
        \hline 

        Total-severity-SS & $2$ & $38$ & $0.31$ & $0.74$ & $0.02$ & $20.01$  & $35.71$ & $11.22$ & $17.58$ & $17.2$ & $32.7$          \\
        \hline 

        Overall Taskload &  \cellcolor{lightgray}$2$  & \cellcolor{lightgray}$38$   & \cellcolor{lightgray}$15.53$   & \cellcolor{lightgray}$<0.001$   & \cellcolor{lightgray}$0.45$ & $34.75$ & $18.16$ & $20.79$ & $13.33$ & $26.38$ & $14.28$         \\
        \hline 
        
        PFA-Finger & \cellcolor{lightgray}$2$ & \cellcolor{lightgray}$38$ & \cellcolor{lightgray}$29.01$ & \cellcolor{lightgray}$<0.001$ & \cellcolor{lightgray}$0.6$  & $2.65$ & $1.14$ & $5.3$ & $1.38$ & $4.25$ & $1.55$         \\
        \hline
        PFA-Control & \cellcolor{lightgray}$2$ & \cellcolor{lightgray}$38$ & \cellcolor{lightgray}$17.13$ & \cellcolor{lightgray}$<0.001$ & \cellcolor{lightgray}$0.47$ & $2.6$ & $1.31$ & $5.1$ & $1.59$ & $3.8$ & $1.51$          \\
        \hline
         PFA-Accuracy & \cellcolor{lightgray}$2$ & \cellcolor{lightgray}$38$ & \cellcolor{lightgray}$14.18$ & \cellcolor{lightgray}$<0.001$ & \cellcolor{lightgray}$0.43$ & $2.9$ & $1.62$ & $5.05$ & $1.47$ & $3.85$ & $1.31$          \\
        \hline
        
    \end{tabular}

    \endgroup

    \caption{\small{RM-ANOVA results of the questionnaire data for the VR-HMDs.  Gray rows show significant findings. SUS: System Usability Scale. SS: Simulator Sickness Questionnaire. PFA: Perceived Finger Assessment.  d$f_1$ = d$f_{effect}$ and d$f_2$ = d$f_{error}$.}}
    \label{tab:questionnaires_VR}
\end{table*}

\subsubsection{Preferences and Open Comments}
We solely asked participants for preferences regarding the VR hand tracking systems, as no finger visualization was employed in the AR HMDs, and, hence we did not expect feedback specific to their hand tracking capabilities. 
From the 20 participants, 17 preferred the \textsc{LeapMotion}. Ten participants said "It was accurate." (P2, P4, P7, P10, P13, P15, P16, P17, P18, P19) and 8 said "It followed my finger movements" (P3, P7, P8, P11, P12, P17, P18, P19). \hl{This is in line with the usability results.}
Three participants preferred the \textsc{Quest}. 
Twelve participants stated the \textsc{Vive} to be the worst HMD in regards of hand tracking accuracy. \hl{This is supported by all measures (DTT, DFT, Z, AFT and RDFT), as the \textsc{VIVE} was significantly less accurate than both \textsc{Quest} and \textsc{LeapMotion}.} 
Six participants said about the \textsc{Vive} that "The tracking indicator was not on my finger" (P5, P11, P15, P16, P17, P18). Five participants said that the \textsc{Quest} was the worst HMD. Three of the participants using the \textsc{Quest} said "The tacking indicator jitters too much" (P8, P12, P13). Three participants could not decide which HMD was the best or worst. Participants were asked to make open comments during the experiment. On the \textsc{LeapMotion}, participants commented: "It feels very responsive."(P03), "much better than before [Quest and Vive] and less frustrating" (P08), "The sphere [fingertip visualization] does not want to go where I want it to" (P06), "It works good while hovering. But on contact with the screen it moves behind it" (P09). On the \textsc{Vive} participants commented: "It feels very accurate but slow" (P04), "I can get used to this" (P09), "The sphere [fingertip visualization] dips too far in the surface of the screen." (P14) and "I hope this won't be used in medical applications" (P12). Comments on the \textsc{Quest} were the following: "It feels better at 0 degrees [\textsc{Horizontal} orientation] " (P13), "Drawing the line with this one is much better" (P19), "Counterclockwise is worse than clockwise [tracing]" (P10) as well as "Oh god, the sphere [fingertip visualization] jitters" (P18).


\section{Robot Study}
While the study with human participants was a controlled experiment and HMDs are likely designed for coping with the hand movements of real users, we also collected measurements using a robotic arm following prior work \cite{weichert2013analysis}. This allows predefined spatial positions to be repeatedly selected with higher accuracy, thereby facilitating reproducibility of results. Please note the trade-off between reproducibility and ecological validity (as no actual human hand but an artificial hand was used). Hence, the results reported in this section should be seen as complementary to the results from the study with human participants.

\subsection{Apparatus}

The setup is shown in Figure \ref{fig:teaser}(f) (for a schematic view see the appendix) and consists of  the robot arm with an artificial hand model (made of silicon) and calibration tool attached and a Styrofoam dummy-head for mounting the HMDs.

The devices and the versions of the software used in the robot study are the same as in the human participants study (see section \ref{sec:apparatus} for more details). The robot used in the study is the Universal Robot UR3 and was mounted on a aluminum frame with its base at a height of 124 cm. 
The UR3 has a reach of 50 cm and it can move the arm to a specific position with an accuracy of $\pm$0.1~mm. The positioning of the robot arm is done with the software of the robot to ensure the arm is always on the correct position. 
The HMD was positioned around 55 cm away from the middle position of the hand and looked directly at mimicking interaction with vertically mounted surfaces. 
The offset of the index fingertip relative to the top left corner of marker was measured with an Optitrack digitizing probe. 
The index fingertip was used as the reference point since it is most similar with the touch point in the study with human participants. 
To avoid potential interference with the tracking systems, the arm of the robot was covered in black fabric. The background was also covered with black fabric. 
To mimic a user's right arm we used a shirt, with the end of the sleeve attached to the hand model and the rest of the shirt attached on the dummy head. In contrast to the study with human participants, the hand model did not show a dedicated pointing pose using the index finger (as such a hand model was not available). However, the operator made sure that the index finger was used for data collection by monitoring the virtual hand model on an external screen. Similar to the study with human participants, a smartphone was used to trigger the next target. 

\subsection{Procedure}
Following the procedure of the study with human participants, the robot fulfilled the same tasks, (see Section \ref{sec:task}), namely the \textsc{Target Acquisition} task with the nine targets repeated five times and the \textsc{Shape Tracing} task with tracing the line and circle again repeated five times, but only in the \textsc{Vertical} orientation and without an actual touch by the silicon hand. For the \textsc{Target Acquisition Task} the robot arm was moved to the specific position and at that position the data was collected. For the \textsc{Shape Tracing Task} the robot arm was moved to the starting position, the data collection was started and then the robot arm moved along the corresponding path. 
In the robot study we only conducted the task in the \textsc{Vertical} orientation as it was not possible for us to mount the arm in a horizontal fashion. We did not use a touchscreen because the touch of the silicon hand could be registered by the touch monitor. 


For each device, the robot study started with a calibration sequence. For this, the robot was moved to the top left position and a calibration tool was used to position the target finger positions (similar to calibrating the monitor in the study with human participants). For the AR devices, the image marker was used and after calibration the marker was covered with black fabric.
For the \textsc{Vive} and the \textsc{Leap Motion}, the Vive Tracker was used for calibrating the target positions. 
After calibration the tracking of the Vive Tracker was disabled in software. For the \textsc{Quest}, the calibration was conducted via the corners of the image marker using the Quest calibration tool. For the \textsc{Target Acquisition} task the robot arm was positioned for the specific target. Then the operator touched on the tablet and collected 3000 samples per target location. After completing the data collection part, the operator triggered the transition to the next target. The operator then again moved the robot arm to the next target position and repeated the process.
For the \textsc{Shape Tracing} task the operator positioned the robot arm at the starting position of the shape and then triggered a prerecorded motion pattern (line or circle). 

The HMDs were not moved on the Styrofoam dummy-head during the data collection process to create a stable environment in contrast to the human participants study, where the user had the freedom to move their head. 
The preparation took ten minutes, calibration lasted five minutes and the main study was conducted within 45 minutes. The overall procedure for each HMD lasted 60 minutes.

\subsection{Results}
As for the study with human participants, additional results can be found in the appendix.

\subsubsection{AR}

\begin{figure}[t]
	\centering 
	\includegraphics[width=\columnwidth]{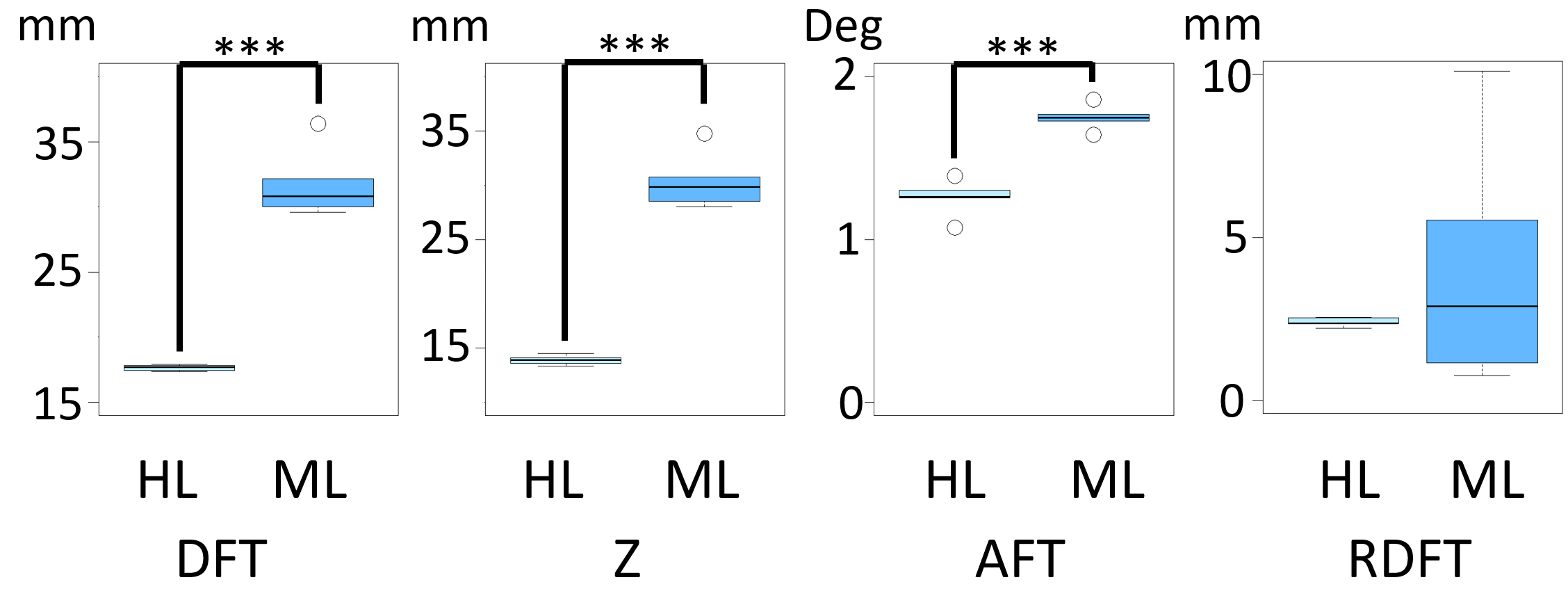}
	\caption{Accuracy measures from the robot study for the \textsc{HoloLens} (HL) and \textsc{MagicLeap} (ML); DFT = Distance Finger-Target in millimeters; Z = Z-Distance in millimeters; AFT = Angle Finger-Target in degrees; RDFT = Radius Difference Finger-Target in millimeters; * = $p<0.05$; ** = $p<0.01$; *** = $p<0.001$.}
	\label{fig:RobotStudy-AR-Boxplot}
\end{figure}

Similar to the human participants study, for the \textsc{Target Acquisition Task} with AR devices, we analyzed the \textsc{Distance Finger-Target} and the \textsc{Z-Distance}. In contrast to the study with human participants, we used paired samples t-test to analyze the data (normality assumptions were met), as there was only one independent variable (\textsc{Interface} with two levels \textsc{HoloLens} and \textsc{MagicLeap}. The results from the analysis and the mean and standard-deviation values are shown in Table \ref{tab:Robot_performance_ratings_AR} and Figure \ref{fig:RobotStudy-AR-Boxplot}. \hl{For the significance tests, the data was log-transformed to ensure a normal-distribution.} The analysis indicates that \textsc{HoloLens} is significantly more accurate than \textsc{MagicLeap} regarding both \textsc{Distance Finger-Target} and \textsc{Z-Distance}.

\begin{table*}[t]
    \centering 
    \small
    \begingroup
    \setlength{\tabcolsep}{5pt}
    
        \begin{tabular}{|>{\centering}m{0.45cm}||c||c||c||c|} 
            \multicolumn{1}{c}{}& \multicolumn{2}{c}{\small\bfseries\textbf{AR - Target Acquisition Task }} & \multicolumn{2}{c}{\small\bfseries\textbf{AR - Shape Tracing Task}} \\
            \hline 
            
            & \multicolumn{1}{c||}{Distance Finger-Target} &\multicolumn{1}{|c||}{Z-Distance} & \multicolumn{1}{c||}{Angle Finger-Target} &\multicolumn{1}{|c|}{Radius Difference Finger-Target}\\
            
            \hline 

            $df$ & \cellcolor{lightgray} $4.0$ & \cellcolor{lightgray} $4.0$ & \cellcolor{lightgray} $4.0$ & $4.0$    \\ 
            
            \hline
            
            $p$ & \cellcolor{lightgray}  $<.001$ & \cellcolor{lightgray} $<.001$ & \cellcolor{lightgray} $<.001$ & $.398$  \\ 

            \hline 

            \tiny{Effect Size} &  \cellcolor{lightgray}  $-6.42$ & \cellcolor{lightgray} $-7.34$ & \cellcolor{lightgray} $-9.45$ & $-0.945$   \\ 

            \hline 

        \end{tabular}
    
        \begin{tabular}{|c|c|c|c|c|c|c|c|c|} 
             \multicolumn{1}{c}{} & \multicolumn{4}{c}{\small\bfseries\textbf{AR - Target Acquisition Task }} & \multicolumn{4}{c}{\small\bfseries\textbf{AR - Shape Tracing Task}} \\
            \hline 
            & \multicolumn{2}{c|}{Distance Finger-Target} & \multicolumn{2}{c|}{Z-Distance} & \multicolumn{2}{c|}{Angle Finger-Target} & \multicolumn{2}{c|}{Radius Difference Finger-Target} \\
            \hline 
            Device & Mean & SD & Mean & SD & Mean & SD & Mean & SD  \\ 
            \hline 

            \textsc{HoloLens} & $17.65~mm$ & $0.24$ & $13.89~mm$  & $0.45$ & $1.26$° & $0.11$ & $2.4$~mm  & $0.14$  \\ 
            
            \hline
            
            \textsc{MagicLeap} & $31.79~mm$ & $2.75$ & $30.39~mm$  & $2.66$ & $1.75$° & $0.08$ & $4.08$~mm  & $3.86$ \\ 

            \hline 

        \end{tabular}

    \endgroup

    \caption{RM-ANOVA results for the robot study and mean and standard-deviation values for AR-devices (regardless of \textsc{Orientation}).  Gray rows show significant findings. I = \textsc{Interface}, O = \textsc{Orientation}. }
    \label{tab:Robot_performance_ratings_AR}
\end{table*}

Similar to the human participants study, for the \textsc{Shape Tracing Task} with AR devices we analyzed the \textsc{Angle Finger-Target} and the \textsc{Radius Difference Finger-Target} (again using a paired samples t-test).
The results from the analysis and the mean and standard-deviation values are shown in Table \ref{tab:Robot_performance_ratings_AR}. \hl{For the significance tests, the \textsc{Angle Finger-Target}-data was log-transformed to ensure a normal-distribution.}  The analysis again indicates that \textsc{HoloLens} is significantly more accurate than \textsc{MagicLeap} regarding \textsc{Angle Finger-Target} but not regarding \textsc{Radius Difference Finger-Target}.

\subsubsection{VR}

Similar to the study with human participants, for the \textsc{Target Acquisition Task} with VR devices we analyzed the \textsc{Distance Finger-Target} and the \textsc{Z-Distance}, but not the \textsc{Distance Touch-Target}, because the touchscreen was not used in the robot study. Because we had one independent variable with three levels, we used a repeated measures ANOVA to analyze the data.
The results from the analysis and the mean and standard-deviation values are shown in Table \ref{tab:robot_performance_ratings_VR}.  \hl{For the significance tests, the data was log-transformed to ensure a normal-distribution.}
The analysis indicates that the \textsc{Interface} significantly influences the \textsc{Distance Finger-Target} and the \textsc{Z-Distance} measures. Post-hoc tests revealed that \textsc{Quest} was significantly more accurate regarding both measures than \textsc{LeapMotion} and \textsc{Vive}. Furthermore, \textsc{LeapMotion} was significantly more accurate than \textsc{Vive}.

\begin{figure}[t]
	\centering 
	\includegraphics[width=\columnwidth]{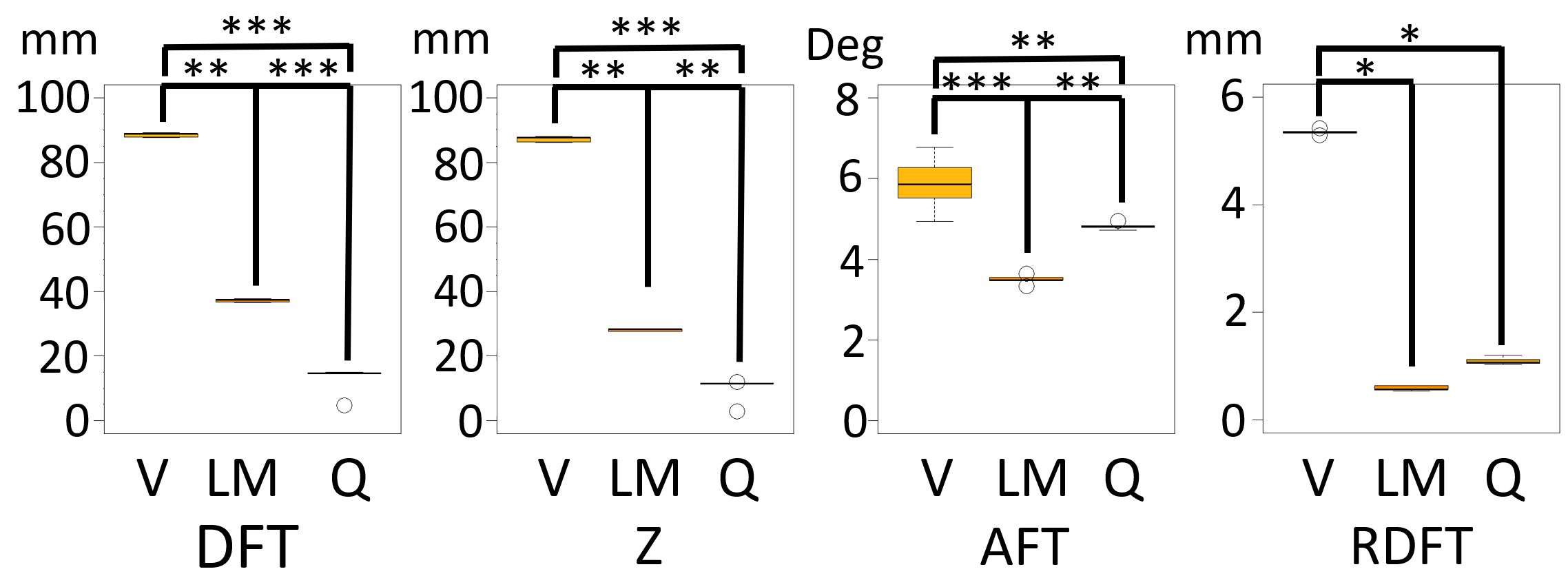}
	\caption{Accuracy measures from robot study for the \textsc{Vive} (V), the \textsc{LeapMotion} (LM) and \textsc{Quest} (Q). DFT = Distance Finger-Target in millimeter; Z = Z-Distance in millimeter; AFT = Angle Finger-Target in degrees; RDFT = Radius Difference Finger-Target in millimeter; * = $p<0.05$; ** = $p<0.01$; *** = $p<0.001$}
	\label{fig:RobotStudy-VR-Boxplot}
\end{figure}

\begin{table*}[t]
    \centering 
    \small
    \begingroup
    \setlength{\tabcolsep}{5pt}

     \begin{tabular}{|>{\centering}m{0.45cm}||>{\centering}m{0.2cm}|>{\centering}m{0.2cm}|>{\centering}m{0.4cm}|>{\centering}m{0.7cm}|>{\centering}m{0.25cm}||>{\centering}m{0.2cm}|>{\centering}m{0.2cm}|>{\centering}m{0.4cm}|>{\centering}m{0.7cm}|>{\centering}m{0.25cm}||>{\centering}m{0.2cm}|>{\centering}m{0.2cm}|>{\centering}m{0.4cm}|>{\centering}m{0.7cm}|>{\centering}m{0.25cm}||>{\centering}m{0.2cm}|>{\centering}m{0.2cm}|>{\centering}m{0.4cm}|>{\centering}m{0.7cm}|c|}  
     
            \multicolumn{1}{c} {} & \multicolumn{10}{c}{\small\bfseries\textbf{VR - Target Acquisition Task }}  &  \multicolumn{10}{c}{\small\bfseries\textbf{VR - Shape Tracing Task }} \\
            \hline 
            & \multicolumn{5}{|c||}{Distance Finger-Target} & \multicolumn{5}{|c||}{Z-Distance} & \multicolumn{5}{c||}{Angle Finger-Target} &\multicolumn{5}{|c|}{Radius Difference Finger-Target} \\
            \hline 
            &   d$f_{1}$ & d$f_{2}$ & F & p &  $\eta^2_p$ & d$f_{1}$ & d$f_{2}$ & F & p &  $\eta^2_p$ & d$f_{1}$ & d$f_{2}$ & F & p &  $\eta^2_p$ & d$f_{1}$ & d$f_{2}$ & F & p &  $\eta^2_p$     \\ 
            \hline 

            I &  \cellcolor{lightgray} $2$ &  \cellcolor{lightgray} $8$ &  \cellcolor{lightgray} $57.0$ &  \cellcolor{lightgray}  $<.001$ &  \cellcolor{lightgray} $0.93$ &  \cellcolor{lightgray} $2$ &  \cellcolor{lightgray} $8$ &  \cellcolor{lightgray} $48.1$ &   \cellcolor{lightgray} $<.001$ &  \cellcolor{lightgray} $0.92$ & \cellcolor{lightgray} $2$ &  \cellcolor{lightgray}$8$ & \cellcolor{lightgray} $43.7$ &  \cellcolor{lightgray} $<.001$ & \cellcolor{lightgray} $0.92$ &\cellcolor{lightgray} $2$ & \cellcolor{lightgray}$8$ & \cellcolor{lightgray}$6.46$ &  \cellcolor{lightgray}$0.021$ & \cellcolor{lightgray}$0.618$   \\ 
            
            \hline 

        \end{tabular}
        
        \begin{tabular}{|c|c|c|c|c|c|c|c|c|} 
            \multicolumn{1}{c} {} & \multicolumn{4}{c}{\small\bfseries\textbf{VR - Target Acquisition Task}}  &  \multicolumn{4}{c}{\small\bfseries\textbf{VR - Shape Tracing Task }} \\
            \hline 
            &   \multicolumn{2}{|c|}{Distance Finger-Target} & \multicolumn{2}{|c|}{Z-Distance} & \multicolumn{2}{c|}{Angle Finger-Target} & \multicolumn{2}{c|}{Radius Difference Finger-Target}  \\
            \hline 
            Device & Mean & SD & Mean & SD & Mean & SD & Mean & SD    \\ 
            \hline 

            \textsc{vive}  & $88.54$~mm & $0.69$ & $87.2$~mm & $0.81$ & $5.87$° & $0.70$ & $5.35$~mm & $0.05$ \\ 
            
            \hline
            
           \textsc{LeapMotion}  & $37.25$~mm & $0.52$ & $28.07$~mm & $0.41$ & $3.5$° & $0.11$ & $3.16$~mm & $5.72$ \\ 

            \hline 

           \textsc{Quest}  & $12.75$~mm & $4.52$ & $9.87$~mm & $3.95$ & $4.82$° & $0.09$ & $1.1$~mm & $0.07$  \\ 

            \hline 

        \end{tabular}
    
    \endgroup

    \caption{RM-ANOVA results for the robot study and mean and standard-deviation values for VR-devices (regardless of \textsc{Orientation}).  Gray rows show significant findings. I = \textsc{Interface}, O = \textsc{Orientation}. }
    \label{tab:robot_performance_ratings_VR}
\end{table*}

Similar to the study with human participants, for the \textsc{Shape Tracing Task} with VR devices we analyzed the \textsc{Angle Finger-Target} and the \textsc{Radius Difference Finger-Target}. As with the \textsc{Target Acquisition Task} we used a repeated measures ANOVA to analyze the data.  The results from the analysis and the mean and standard-deviation values are shown in Table \ref{tab:robot_performance_ratings_VR}. \hl{For the significance tests, the \textsc{Radius Difference Finger-Target}-data was log-transformed to ensure a normal-distribution.}
The analysis indicates that the \textsc{Interface} significantly influences the \textsc{Angle Finger-Target} and the \textsc{Radius Difference Finger-Target}. Post-hoc tests indicate that the angle between the target line and the line through the tracked finger-points was significantly smaller for \textsc{LeapMotion} compared to\textsc{Vive} and \textsc{Quest} and that it was significantly smaller for \textsc{QUEST} compared to \textsc{Vive}. The \textsc{Vive} differed significantly more from the target shapes than \textsc{LeapMotion} and \textsc{Quest}.






\section{Discussion}
Our study reported on the accuracy of AR and VR commodity HMDs for finger tracking when interacting with surfaces. 
Our findings suggest that for VR HMDS, the HTC Vive results in significantly lower spatial accuracy compared to both Oculus Quest and Leap Motion. For AR HMDs the Magic Leap One resulted in significantly lower spatial accuracy compared to the Microsoft HoloLens 2. 
Those results were indicated both in the study with human participants and in the robot study. 
\hl{They are also supported by the subjective feedback of the participants, as most participants (17 out of 20) preferred the Leap Motion and more than half (12/20) liked the HTC Vive least.}

The indicated spatial accuracy results have to be understood as a cumulative measure, integrating multiple tracking systems. Specifically, the accuracy of the respective HMD tracking system is added to the accuracy of the actual sensors employed for hand and finger tracking. 

In contrast to prior work, the mean spatial accuracy results reported in this paper are typically lower (specifically for the Leap Motion). For example while Weichert et al. \cite{weichert2013analysis} reported spatial accuracy of below 0.2 mm for the Leap Motion controller, in our test setup the mean spatial accuracy in a pointing task was 24 mm (sd = 8) for the study with human participants. The difference can potentially be attributed to different sensor-hand configurations (e.g., sensor lying on a table with the hand interacting up close vs.~head-mounted sensor with hand interacting at arm's reach). Also, we witnessed a high inter-subject variability of the hand tracking accuracy. 

For the AR HMDs and the \textsc{QUEST}, the robot study mostly replicated the results from the human participants study. However, for the \textsc{Vive} and \textsc{LeapMotion}, the mean spatial accuracy for the target acquisition task resulted in approximately three times higher mean distances in the robot study compared to the human participants study. We see multiple possible explanations for this diverging behaviour: First, in the VR conditions, users typically aligned their virtual finger tip with the respective target, potentially lowering the accuracy error compared to when they would align their physical fingertip with the target. In contrast, in the robot study no such compensating movement was made, but the physical fingertip of the artificial hand was aligned with the target. Second, in the study with human participants, the participants moved their hand back and forth in space when switching between the homing position and the touchscreen. This potentially allowed the tracking systems to get a better initial estimate of the finger position which could then be updated as the finger moved towards the touch surface. Also, in the study with human participants, participants were free to move their head back and forth to facilitate tracking. In contrast, in the robot study, the artificial hand was moved in a fixed plane (mostly) parallel to the HMD. Also, the HMD was fixed. This resulted in substantially less variance in the Z-distance between the hand and the HMD. Third, compared to the Oculus Quest, \textsc{Vive} and \textsc{LeapMotion} have substantially smaller baselines for their sensors (\textsc{Vive} uses two cameras with a baseline of ca. 6.5~cm, \textsc{LeapMotion} also uses two cameras with a baseline of ca. 4 cm, Quest uses four cameras with horizontal baselines between 12 and 15~cm and vertical baselines of approximately 7.5~cm). This also likely contributes to a worse accuracy at larger distances for \textsc{Vive} and \textsc{LeapMotion} compared to \textsc{Quest}.




The paper is looking only at tracking a single fingertip, visible to the tracking cameras. However, there are additional factors that may reduce the quality of tracking. For example, occlusion of the fingers by the hand or another hand or object may increase the error of fingertips locations. While the general shape of the hand, as rendered by the HMD, may look plausible, location-critical rendering, such as use of world-grounded haptics may be sensitive to such errors. Also, fast movements of the hands or the head may reduce the quality of tracking due to motion blur. In this paper we focused on moderate hand motions and head rotations. Finally, there are hand trackers, such as the Leap Motion, that may be positioned on the screen, or on the table. Due to being separated from the HMD, this generates problems of portability, power, and calibration, but these locations may give the sensors better view of the fingers, unclouded by the back of the hand, and less dependency on the accuracy of the head tracking by the HMD.



\subsection{Limitations}
Similar to prior work \cite{xiao2018mrtouch}, our study focused on accuracy as the primary objective measure for characterising performance for touch-based interaction. We deliberately chose this measure to be able to compare measures from the study with human participants and the robot study. However, prior work \cite{teather2009effects} has indicated that latency can have a stronger effect on human performance in 2D pointing and constrained 3D object movement tasks. Hence future work may consider exploring throughput as an additional measure for jointly capturing movement speed and accuracy in pointing tasks. Also, our work focused on the combination of a VR HMD with the Leap Motion Sensor, but we are aware that the Leap Motion is also used in AR HMDs (e.g., \cite{batmaz2020touch}). However, prior work also suggests that differences in performance between AR and VR HMDS when using the same sensing system, such as Leap Motion, is likely due to the display system \cite{batmaz2020touch}. For the AR HMDs we deliberately did not show a virtual finger tip to avoid confusion on whether to align the physical or virtual finger with the target. However, showing a virtual finger tip could allow users to better react to potential tracking issues, and, therefore lead to a higher spatial accuracy. In the study with human participants we used both \textsc{vertical} and \textsc{horizontal} orientations while in the robot study we only used the \textsc{vertical} orientation, as it was not possible for us to mount the arm in a horizontal fashion. We also did not use a touchscreen in the robot study since the robot could position the hand accurately in a predefined plane. Hence, technically, no touch interactions were performed in the robot study. Also, the lack of a touchscreen could potentially be considered a confounding factor as potentially distracting reflections of the touch display were not present in the robot study. We also considered including further orientations (e.g., 45$^{\circ}$), but did not do so due to the already long duration of the study with human participants. Finally, the study solely reported results from the overall spatial accuracy. Theoretically, individual components of the HMDs overall tracking architecture could be exchanged, for example, the hand tracking systems could be combined with outside-in tracking systems for tracking the HMD position.

\section{Conclusions}

We have evaluated commodity hand tracking systems for AR and VR HMDs including the Oculus Quest, Vive Pro, Microsoft HoloLens 2 and Magic Leap One, as well as the Leap Motion controller. We reported and compared the accuracy for absolute and relative pointing tasks in order to inform researchers' and designers' decisions when building new systems or experiences using finger tracking.
Future work includes user experience research with touch displays and finger tracking to allow supporting touch and hover tracking at the same time to create experiences that uses the approach trajectory of the fingers to seamlessly adapt the user interface accordingly. The accuracy of hand tracking when interacting in mid-air is also a fruitful avenue of future work. To this end, controlled experiments could be conducted investigating performance in larger interaction volumes.

\balance
\bibliographystyle{ACM-Reference-Format}
\bibliography{acmart}


\end{document}